\documentclass[10pt,twoside]{epnt1p}%
\usepackage{amsmath}
\usepackage{amsfonts}
\usepackage{amssymb}
\begin{document}

\begin{frontmatter}
\title{ Strangelets, Nuclearites, Q-balls---A Brief Overview}
\author{Jes Madsen}
\address{Department of Physics and Astronomy, University of Aarhus, Denmark }
\begin{abstract}
Astrophysical bounds on the properties and abundances of primordial quark nuggets
and cosmic ray strangelets are reviewed. New experiments to search for cosmic
ray strangelets in lunar soil and from the International Space Station
are described. Analogies with baryonic and supersymmetric
Q-balls are briefly mentioned, as are prospects for strangelets as ultra-high energy
cosmic rays.
\end{abstract}
\end{frontmatter}


\section{ Introduction}

Quark nuggets, nuclearites and strangelets are different names for lumps of a
hypothetical phase of absolutely stable quark matter, socalled strange quark
matter because of the admixture of slightly less than one-third strange quarks
with the up and down quarks. Whether strange quark matter is absolutely stable
is a question yet to be decided by experiment or astrophysical observation
(see \cite{Weber:2004kj,Madsen:1998uh} for reviews), 
but if it is the case, then strange quark matter
objects may exist with baryon numbers ranging from ordinary nuclei to a
maximum of order $2\times10^{57}$ corresponding to gravitational instability
of strange stars.

Truly macroscopic quark matter lumps surviving from the cosmological
quark-hadron phase transition are often referred to as quark nuggets, and if
they hit the Earth they are sometimes dubbed nuclearites. Strangelets are
smaller lumps (baryon number $A<10^{7}$) 
where the electron cloud neutralizing the slightly
positive quark charge mainly resides outside the quark core. Strangelets are
unlikely to survive from the early Universe, but may form as a result of strange
star binary collisions and/or acceleration from the surface of pulsars.
The nomenclature is not strictly defined, and in the following the word
strangelet will be used as a general name except when discussing leftovers
from the early Universe.

Q-balls are non-topological solitons suggested from various origins in the
early Universe and as such have nothing to do with strange quark matter (their
origin and general properties were described by Kusenko at this workshop).
However for some classes of Q-balls (baryonic and supersymmetric Q-balls), the
astrophysical bounds that can be derived on strangelets are easily generalized
to these creatures as well and are therefore of interest in the context of
this Workshop on Extreme Physics with Neutrino Telescopes.

In the following I shall discuss the (unlikely) survival of cosmological quark
nuggets, the more optimistic prospects for cosmic rays strangelets from
strange stars, and two new experimental efforts to search for them. Due to
space limitations I will not go into details with the Q-ball analogies but
refer the reader to Kusenko's contribution and to \cite{madqball}.

\section{Sources, sizes, and fluxes}

\subsection{Primordial quark nuggets}

In a first order cosmological quark-hadron phase transition at $T\approx
100\,$MeV, supercooling may lead to concentration of baryon number inside
shrinking bubbles of quark phase, that may reach nuclear matter density and
form quark nuggets. The baryon number inside the horizon during the cosmic
quark-hadron phase transition (an upper limit for causal formation of quark
nuggets) is $A_{\mathrm{hor}}\approx10^{49}.$ Witten \cite{wit84} predicted
that typical nuggets would be somewhat smaller than this and argued that quark
nuggets might explain the cosmological dark matter problem. Quark nuggets
would decouple from the radiation bath very early in the history of the
Universe and behave as cold dark matter in the context of galaxy formation.
Today the nuggets would move with typical galactic halo velocities of a few
hundred kilometers per second through the Milky Way.

Later studies showed that the hot environment made cosmological nuggets
unstable against surface evaporation \cite{alcfar85}
and boiling \cite{alcoli89}, effectively destroying nuggets
with $A$ below $10^{39-46}$ depending on assumptions. Small traces of
primordial nuggets with lower baryon numbers could also be left over from the
destruction processes in the early Universe. Even such traces may in fact be
\textquotedblleft observed\textquotedblright\ using the astrophysical
detectors discussed below. Let $v\equiv250\mathrm{km\,s}^{-1}v_{250}$ and
$\rho\equiv10^{-24}\mathrm{g\,cm}^{-3}\rho_{24}$ be the typical speed and mass
density of nuggets in the galactic halo. The speed is given by the depth of
the gravitational potential of our galaxy, whereas $\rho_{24}\approx1$
corresponds to the density of dark matter. Then the number of nuggets hitting
the Earth is
\begin{equation}
\mathcal{F}\approx 6.0\times10^{5}A^{-1}\rho_{24}v_{250}\,\mathrm{cm}^{-2}%
\,\mathrm{s}^{-1}\,\mathrm{sr}^{-1}.
\end{equation}

Quark nuggets have a positive electrostatic surface potential (several MeV) of
the quark phase because quarks are stronger bound than electrons, so
during Big Bang nucleosynthesis ($T\leq1$ MeV), nuggets absorb neutrons but
not protons. This reduces the neutron-to-proton ratio, thereby lowering the
production of $^{4}$He. The helium-production is very sensitive to the total
amount of nugget-area present, and in order not to ruin the concordance with
observations, one finds \cite{madrii85} that only nuggets with
$A>A_{\mathrm{BBN}}\approx10^{22}\Omega_{\mathrm{nug}}^{3}f_{n}^{3}$ are
allowed during nucleosynthesis. Here $\Omega_{\mathrm{nug}}$ is the
present-day nugget contribution to the cosmic density in units of the critical
density, and $f_{n}\leq1$ is the penetrability of the nugget surface. 

In spite
of carrying baryon number, primordial quark nuggets do \textit{not\/}
contribute to the usual nucleosynthesis limit on $\Omega_{\mathrm{baryon}}$.
The baryon number is \textquotedblleft hidden\textquotedblright\ in quark
nuggets long before Big Bang nucleosynthesis begins, and the nuggets only
influence nucleosynthesis via the neutron absorption just described.
The same is true for baryon number carrying Q-balls.

While primordial quark nuggets remain a possibility within the 
tight restrictions
mentioned, the main problem with this scenario is the need for a first order
quark-hadron phase transition which is currently not favored in lattice QCD
studies at zero chemical potential.

\subsection{Strangelets from compact stars}

If strange quark matter is absolutely stable all compact stars are likely to
be strange stars (see the following Section), and therefore the galactic
coalescence rate estimated for neutron star binaries that inspiral due to loss
of orbital energy by emission of gravitational radiation, believed to be of
order one collision in our Galaxy every 10,000 years
\cite{Kalogera:2003tn}, is really the rate of
strange star collisions. Each event involves a phase of tidal disruption as
the stars approach each other before the final collision. During this stage
small fractions of the total mass may be released from the binary system in
the form of strange quark matter. No realistic simulation of collisions
involving two strange stars has been performed. Simulations of binary neutron
star collisions, depending on orbital and other parameters, lead to the
release of anywhere from $10^{-5}-10^{-2}M_{\odot}$
($M_{\odot}$ is the solar mass), corresponding to a total
mass release in the Galaxy of $10^{-10}-10^{-6}M_{\odot}$ per year. The
equation of state for strange quark matter is stiff, so strange star
collisions probably lie in the low end of the mass release range. A
conservative estimate of the galactic production rate of strangelets is
$10^{-10}M_{\odot}$yr$^{-1}.$

Quark matter lumps released by tidal forces are macroscopic 
\cite{Madsen:2001bw},
but subsequent collisions lead to fragmentation, and if the collision energy
compensates for the surface energy involved in making smaller strangelets, a
significant fraction of the mass released from binary strange star collisions
might end up in the form of strangelets with $A\approx10^{2}-10^{4}$
\cite{Madsen:2001bw}.

Incidentally, a similar range of strangelet masses is expected if the surface
of strange stars consists of a layer with strangelets embedded in an electron
gas rather than pure quark matter all the way to the surface 
\cite{Jaikumar:2005ne}.

Assuming that strangelets from binary collisions are accelerated
and propagate like cosmic ray nuclei in our Galaxy, taking
proper account of their small charge-to-mass ratio, as well as energy loss,
spallation, escape from the Galaxy, etc., it was shown in 
\cite{Madsen:2004vw} that the
expected flux for color-flavor locked strangelets (charge $Z=0.3A^{2/3}$
\cite{Madsen:2001fu}) near Earth is
\begin{equation}
\mathcal{F}\approx10^{-6}A^{-1.47}\mathrm{cm}^{-2}\,\mathrm{s}^{-1}%
\,\mathrm{sr}^{-1}.
\label{appflux}
\end{equation}
Most of these nuggets have rigidities (momentum divided by charge) of a few
GV, but with a powerlaw tail at higher rigidity. Apart from the slightly
different $A$-dependence this is some 12 orders of magnitude smaller than the
flux estimate for dark matter nuggets, which is not unreasonable because the
total strangelet mass originating from binary collisions over the age of the
Galaxy is around 1 $M_{\odot}$, compared to $10^{12}$ $M_{\odot}$ of dark matter.

Another possible cosmic ray strangelet source is extraction from the
surface of pulsars and acceleration in the strong pulsar electric
fields. A measurable flux is predicted in \cite{Cheng:2006ak} 
within the scenario where the strange star surface consists of
strangelets embedded in an electron gas \cite{Jaikumar:2005ne}.
Formation and acceleration in
supernova explosions has also been suggested \cite{Benvenuto:1989hw}.

\section{Detection of cosmic ray strangelets (or why either all or no compact
stars are strange)}

De R\'{u}jula and Glashow \cite{dergla84} argued that quark nuggets hitting
the Earth would show up as unusual meteor-events, earth-quakes, etched tracks
in old mica, in meteorites and in cosmic-ray detectors. A negative search for
tracks in ancient mica corresponded to a lower nugget flux limit of
$8\times10^{-19}\,\mathrm{cm}^{-2}\,\mathrm{s}^{-1}\,\mathrm{sr}^{-1}$ for
nuggets with $A>1.4\times10^{14}$ (smaller nuggets being trapped in layers
above the mica samples studied).

Later investigations have improved these flux limits by a few orders of
magnitude and extended them to lower $A$, though with higher flux limits (see
other contributions to these proceedings for examples). This has excluded
quark nuggets with $3\times10^{7}<A<5\times10^{25}$ as dark matter, but a low
flux from the Big Bang or from collision of strange stars cannot be ruled out.

Neutron stars and their stellar progenitors\ may be thought of
as alternative large
surface area, long integration time detectors leading to much tighter flux
limits \cite{mad88}, see also \cite{calfri91}. 
The presence of a single quark nugget in
the interior of a neutron star is sufficient to initiate a transformation of
the star into a strange star \cite{wit84,alcfar86a}. The time-scale
for the transformation is short, between seconds and minutes, so observed
pulsars would have been converted long ago if their stellar progenitors ever
captured a quark nugget, or if neutron stars absorbed one after formation.

To convert a neutron star into strange matter a quark nugget should not only
hit a supernova progenitor but also be caught in the core \cite{mad88}. A main
sequence star is capable of capturing non-relativistic quark nuggets with
baryon numbers below $A_{\mathrm{STOP}}\approx10^{31}$, which are braked by
inertia, i.\ e.\ they are slowed down by electrostatic scatterings after
plowing through a column of mass similar to their own, and settle in the
stellar core. Relativistic nuggets may be destroyed after collisions with
nuclei in the star, but even a
tiny fraction of a nugget surviving such an event and settling in the star is
sufficient to convert the neutron star to a strange star, so the
non-relativistic flux limits may still apply.

For nuggets with $A<A_{\mathrm{STOP}}$ the sensitivity of main sequence stars
as detectors is given by the limit of one nugget hitting the surface of the
supernova progenitor in its main sequence lifetime. Converted into a flux,
$\mathcal{F}$, of nuggets hitting the Earth it corresponds to
\begin{equation}
\mathcal{F}\approx 4\times10^{-42}v_{250}^{2}\mathrm{cm}^{-2}\,\mathrm{s}%
^{-1}\,\mathrm{sr}^{-1}.
\end{equation}
This is a factor of $10^{20}$--$10^{40}$ more sensitive 
than direct detection experiments!

\textit{If} it is possible to prove that some neutron stars are indeed neutron
stars rather than strange stars, the sensitivity of the astrophysical
detectors rules out quark nuggets as dark matter for $A<10^{34-38}$. And it
questions the whole idea of stable strange quark matter, since it is
impossible to avoid polluting the interstellar medium with nuggets from
strange star collisions or supernova explosions at fluxes many orders of
magnitude above the limit measurable in this way.

\textit{If} on the other hand strange quark matter
is stable, then all neutron stars are likely
to be strange stars, again because some pollution can not be avoided.

\section{Experiments underway}

Several experiments have searched for strangelets in cosmic rays. While some
interesting events have been found that are consistent with the predictions
for strangelets, none of these have been claimed as real discoveries.
Interpreted as flux limits rather than detections these results are consistent
with the flux estimates given above. For discussions see
\cite{Sandweiss:2004bu,Finch:2006pq}.

If the interesting events were actual measurements, two experiments that are
currently underway will reach sensitivities that would provide real statistics.

\textbf{AMS-02: }The Alpha Magnetic Spectrometer (AMS) is a space-based
particle physics experiment involving several hundred physicists from more
than 50 institutions in 16 countries, led by Nobel laureate Samuel Ting of
MIT. A prototype (AMS-01) flew in June 1998 aboard the Space Shuttle Discovery
\cite{Aguilar:2002ad}, and AMS-02 is currently scheduled to fly to the
International Space Station (ISS) in 2009. Once on the ISS AMS-02 will remain
active for at least three years. AMS-02 will provide data with unprecedented
accuracy on cosmic ray electrons, positrons, protons, nuclei, anti-nuclei and
gammas in the GV-TV range and probe issues such as antimatter, dark matter,
cosmic ray formation and propagation. In addition it will be uniquely suited
to discover strangelets characterized by extreme rigidities for a given
velocity compared to nuclei \cite{Sandweiss:2004bu,Finch:2006pq}. 
AMS-02 will have
excellent charge resolution up to $Z\approx30$, and should be able to probe a
large mass range for strangelets. A reanalysis of data from the AMS-01
mission has given hints of some interesting events, such as one with
$Z=2,A=16$ \cite{Choutko:2003} and another with $Z=8,$ but with the larger
AMS-02 detector running for 3 years or more, real statistics is achievable.

\textbf{LSSS: }The Lunar Soil Strangelet Search (LSSS) is a search for $Z=8$
strangelets using the tandem accelerator at the Wright Nuclear Structure
Laboratory at Yale \cite{Finch:2006pq,Han:2006}. 
The experiment involves a dozen people from Yale, MIT, and
\AA rhus, led by Jack Sandweiss of Yale. The experiment which is about to
begin its real data taking phase, studies a sample of 15 grams of lunar soil
from Apollo 11. It will reach a sensitivity of $10^{-17}$ over a wide mass
range, sufficient to provide detection according to Eq.~\eqref{appflux}
if strangelets have been trapped in the lunar surface layer, which has an
effective cosmic ray exposure time of around 500 million years and an
effective mixing depth due to micrometeorite impacts of only around one
meter, in contrast to the deep geological and oceanic mixing on Earth.
Combined with the fact that the Moon has no shielding magnetic field,
this results in an expected strangelet concentration in lunar soil
which is at least four orders of magnitude larger than the corresponding
concentration on Earth.

\section{Strangelets as ultra-high energy cosmic rays}

The existence of cosmic rays with energies well beyond $10^{19}\mathrm{eV}$,
with measured energies as high as $3\times10^{20}\mathrm{eV}$, is one of the
most interesting puzzles in cosmic ray physics
\cite{Greisen:1966jv}. It is almost impossible to find a
mechanism to accelerate cosmic rays to these energies. Furthermore ultra-high
energy cosmic rays lose energy in interactions with cosmic microwave
background photons, and only cosmic rays from nearby (unidentified) sources
would be able to reach us with the energies measured. Strangelets circumvent
both problems \cite{Madsen:2002iw},
and therefore provide a possible mechanism for cosmic rays
beyond the socalled GZK-cutoff.

\textbf{Acceleration: }All astrophysical \textquotedblleft
accelerators\textquotedblright\ involve electromagnetic fields, and the
maximal energy of a charged particle is proportional to its charge. The charge
of massive strangelets has no upper bound in contrast to nuclei, so highly
charged strangelets are capable of reaching energies much higher than those of
cosmic ray protons or nuclei using the same \textquotedblleft
accelerator\textquotedblright\ \cite{Madsen:2002iw}.

\textbf{The GZK-cutoff} is a consequence of ultrarelativistic cosmic rays
hitting a $2.7\mathrm{K}$ background photon with a Lorentz-factor $\gamma$
large enough to boost the $7\times10^{-4}\,\mathrm{eV}$ photon to energies
beyond the threshold of energy loss processes, such as photo-pion production
or photo-disintegration. The threshold for such a process has a fixed energy,
$E_{\mathrm{Thr}}$, in the frame of the cosmic ray, e.g., $E_{\mathrm{Thr}%
}\approx10\mathrm{MeV}$ for photo-disintegration of a nucleus or a strangelet,
corresponding to $\gamma_{\mathrm{Thr}}=E_{\mathrm{Thr}}/E_{2.7\mathrm{K}%
}\approx10^{10},$ or a cosmic ray total energy
\begin{equation}
E_{\mathrm{Total}}=\gamma_{\mathrm{Thr}}Am_{0}c^{2}\approx10^{19}%
A~\mathrm{eV.}%
\end{equation}
Since strangelets can have much higher $A$-values than nuclei, this pushes the
GZK-cutoff energy well beyond the current observational limits for ultra-high
energy cosmic rays \cite{Madsen:2002iw,Rybczynski:2001bw}.

\section{Conclusion}

Lumps of strange quark matter (quark nuggets, nuclearites, strangelets) may
form in a first-order cosmological quark-hadron phase transition (unlikely),
or in processes related to compact stars (more likely). Flux estimates for
lumps reaching our neighborhood of the Galaxy as cosmic rays are in a range
that makes it realistic to either detect them in upcoming experiments like
AMS-02 or LSSS, or place severe limits on the existence of stable strange
quark matter. A similar line of reasoning applies to Q-balls.

\section*{Acknowledgments}

{\small This work was supported by the Danish Natural Science Research
Council. }

\setcounter{section}{0} \setcounter{subsection}{0} \setcounter{figure}{0}
\setcounter{table}{0} \newpage

\end{document}